\documentclass[article, groupedaddress, 11pt]{revtex4}

\usepackage{amsmath,amssymb,amsfonts, url, subfigure, dsfont, mathrsfs, graphicx}
\newcommand{\ds}{\displaystyle}
\newtheorem{theo}{Theorem}[section]

\long\def\symbolfootnote[#1]#2{\begingroup%
\def\thefootnote{\fnsymbol{footnote}}\footnote[#1]{#2}\endgroup}
\input epsf
\begin{document}

\title{Recovering the time-dependent transmission rate from infection data via solution of an inverse ODE problem}
\author{Mark Pollicott$^1$, Hao Wang$^{2\dag}$, and Howie Weiss$^3$} \affiliation{$^1$
Mathematics Institute, University of Warwick, Coventry, CV4 7AL,
United Kingdom
\\ $^2$Department of Mathematical and Statistical Sciences, University of Alberta,
Edmonton, Alberta, T6G 2G1, Canada
\\ $^3$School of Mathematics, Georgia Institute of Technology, Atlanta, Georgia, 30332, United States
\\ $\dag$ To whom correspondence should be addressed. Phone: 1-780-492-8472. Fax: 1-780-492-6826. Electronic address: hwang@math.ualberta.ca}

\begin{abstract}
\vspace{4ex}
\begin{center}\textbf{Abstract}\end{center}
The transmission rate of many acute infectious diseases varies
significantly in time, but the underlying mechanisms are usually
uncertain. They may include seasonal changes in the environment,
contact rate, immune system response, etc. The transmission rate has
been thought difficult to measure directly. We present a new
algorithm to compute the time-dependent transmission rate directly
from prevalence data, which makes no assumptions about the number of
susceptibles or vital rates. The algorithm follows our complete and
explicit solution of a mathematical inverse problem for SIR-type
transmission models. We prove that almost any infection profile can
be perfectly fitted by an SIR model with variable transmission rate.
This clearly shows a serious danger of over-fitting such
transmission models. We illustrate the algorithm with historic UK
measles data and our observations support the common belief that
measles transmission was predominantly driven by school contacts.

\bigskip


\end{abstract}

\maketitle

\pagestyle{myheadings} \thispagestyle{plain} \markright{Recovery
algorithm}
\newpage

\section{Introduction}
%{\bf The SIR epidemic model was proposed by \citet{ker27} and was
%extensively developed by \citet{and92}. One of the key parameters in these transmission models is
%the transmission rate, which is the product of the average number of
%contacts a susceptible individual has with infected individuals per
%unit time and the average probability of transmission during each effective contact.}

The transmission rate of an infectious disease is the rate of which
susceptible individuals become infected. In Section 3.4.9 of
\citet{and92}, the authors state that ``{\it ... the direct
measurement of the transmission rate  is essentially impossible for
most infections. But if we wish to predict the changes wrought by
public health programmes, we need to know the transmission rate ...
.'' } The transmission rate of many acute infectious diseases varies
significantly in time and  frequently exhibits  significant seasonal
dependence  \citep{bec99,scott,wol08}: influenza, pneumococcus, and
rotavirus cases peak in winter; respiratory syncytial virus and measles cases peak in
spring; and polio cases peak in summer.

%Measles outbreaks in several UK cities (pre-vaccination) exhibited  an intriguing
%two-year cycle (see Figure 3(a)(b)), and many investigators have
%attempted to explain this two-year cycle using mathematical models.
%Since measles cases, as well as cases of several other childhood
%viral diseases such as rubella and pertussis, seem to strongly
%correlate with school terms and breaks, most previous modelers have
%assumed that the measles transmission rate  can be represented by a
%term-time forcing function (such as sinusoidal \citep{grass01} or
%Haar \citep{kel} function) with one-year period corresponding to the
%school year. This assumption is based on little empirical evidence
%and ignores the seasonal changes of the environment, immune system,
%etc. \citep{fuj98,pat00} on the transmission rate. In
%Section II(C) we apply our inverse

%For example,
%since measles spreads through respiration directly or indirectly
%through aerosol, the low absolute humidity indoors and outdoors
%during winter may constrain both transmission efficiency and the
%virus survival rate \citep{rem85,sha09}.

Most investigators \emph{define} the transmission rate $\beta(t)$ of
an infectious disease via the discrete transmission model:
\begin{eqnarray}\label{eqn:xxx}
  S(k+1) &=& S(k)-I(k+1), \\
  I(k+1) &=& \beta(k)S(k)I(k),
\end{eqnarray}
where $S(k), I(k)$ are the fractions of susceptible and infected
individuals during week $k$ \citep{fin82, bec89}. Equation (2) is
equivalent to $\beta(k)=I(k+1)/[I(k)S(k)]$, which provides a formula
for the transmission rate. Application of this ``algorithm'' requires knowledge
of $S(0)$,  which depends on vital rates and in general is believed to be very difficult to estimate. We developed our algorithm, in part,  to avoid having to estimate $S(0)$.
%
%This method was modified by
%\citet{bjo02,fin00} to include removal.

Our  new algorithm is based on the soluition of a mathematical
inverse problem for SIR-type transmission models. We
first consider the simplest SIR model (\citet{ker27})
and allow the transmission rate to be a time-dependent function,
\emph{i.e.}, there is a positive function $\beta(t)$ such that
\begin{eqnarray}\label{equ1}
S'(t) &=& -\beta(t) S(t)I(t), \\\label{equ2} I'(t)&=& \beta(t)
S(t)I(t)-\nu I(t), \\\label{equ3} R'(t) &=& \nu I(t),
\end{eqnarray}
where $S(t), I(t)$, and $R(t)$ are the fractions of susceptible,
infected, and removed individuals at time $t$.
%The original
%transmission term $\beta(t) S(t)I(t)/N$ ($S(t)$-number of
%susceptible individuals, $I(t)$-number of infected individuals) is
%scaled to $\beta(t) S(t)I(t)$ ($S(t)$-proportion of susceptible
%individuals, $I(t)$-proportion of infected individuals) if we divide
%all original equations by the total population $N$.
We  pose the mathematical question:

\textit{ Given ``smooth infection data" $f(t)$ on time interval $[0, T]$
and removal rate $\nu>0$, can one always find  a non-negative transmission rate
function $\beta(t)$ such that the $I(t)$ output of the SIR
model always coincides with $f(t)$ with the given $\nu$?}

Mathematicians call this an inverse problem. We prove that this is
always possible subject to a mild restriction on the infection data
and $\nu$, and we provide an explicit formula for the solution. The
construction also illustrates the danger of overfitting a
transmission model where one can choose the time-dependent
transmission rate.

However, in practice, infection data are always discrete, not
continuous. We show that one can robustly estimate $\beta(t)$ by first
smoothly interpolating the data with a spline or trigonometric function
and then applying the formula to smooth data.

The usual transmission rate recovery method based on (1) and (2) can
be viewed as a discretization of (3) and (4). However, unlike the
discrete recovery method, our method extends to a large spectrum of
transmission models (see Section IV B) including those with
different transmission modes or immunity memory periods. Our
approach also yields an explicit  formula for the
transmission rate.

One of the extensions in Section IV B is to the SEIR epidemic model
with historical (time-dependent) vital rates, and we illustrate this
extended algorithm using UK measles data during 1948-1966.
Our recovered transmission rate exhibits two  dominant
spectral peaks: at frequencies $1$ and $3$ per year, respectively.
We show that the latter peak reflects the ``cycle'' of Christmas, Easter, and Summer  school vacations. These
 observations support  the common belief that measles transmission is
predominantly driven by school contacts \citep{grass01, kel}. However, we also
find indications of a cycle with $2$ year period.

%During this period measles outbreaks in several UK cities exhibited  an intriguing
%two-year cycle (see Figure 3(a)(b)), and many investigators have
%attempted to explain this two-year cycle using mathematical models.
%Since measles cases, as well as cases of several other childhood
%viral diseases such as rubella and pertussis, seem to strongly
%correlate with school terms and breaks, most previous modelers have
%assumed that the measles transmission rate  can be represented by a
%term-time forcing function (such as sinusoidal \citep{grass01} or
%Haar \citep{kel} function) with one-year period corresponding to the
%school year. This assumption is based on little empirical evidence
%and ignores the seasonal changes of the environment, immune system,
%etc. \citep{fuj98,pat00} on the transmission rate. In
%Section II(C) we apply our inverse

%
%To
%estimate the initial transmission rate $\beta(0)$, we numerically
%examine the upper threshold of $\beta(0)$ leading to stationary
%$\beta(t)$. When initial transmission rates are larger than this
%threshold, the recovered transmission rates are non-stationary
%functions - they exhibit an upward drift, for which there is no
%public health or biological justification. When initial transmission
%rates are smaller than this threshold, the recovered transmission
%rates are stationary functions.

\section{Results}

We first derive the  algorithm for recovering the
time-dependent transmission rate from infection data. We then apply
the algorithm to two simulated data sets representing two
characteristic ``types" of infectious diseases. Finally we illustrate
the algorithm using UK measles data from  1948-1966.

\subsection{Solution of inverse problem }

The new  algorithm follows from the complete solution of an  inverse problem for the SIR system of ODEs. The generality of  the result seems striking, while the proof is almost trivial.

\begin{theo}\label{theory2}
Given a smooth positive function $f(t)$, $\nu>0$, $\beta_0> 0$, and $T > 0$, there
exists $K > 0$ such that if $\beta_0 < K$ there is a solution
$\beta(t)$ with $\beta(0)=\beta_0$ such that $I(t)=f(t)$ for $0\leq
t\leq T$  if and only if $f'(t)/f(t) > - \nu $ for  $0 \leq t \leq
T$.
\end{theo}

The
growth condition  imposes no restrictions on
how $f(t)$ increases, but requires that $f(t)$ cannot decrease too
quickly, in the sense that its logarithmic derivative is always bounded
below by $-\nu$. It is easy to see that $f'(t)/f(t) > -\nu$ is a necessary condition,
since Equation (\ref{equ2}) implies that ${f'(t)}+\nu f(t)=\beta(t)
S(t) f(t)$, which must be positive for $0\leq t\leq T$.

 The proof of the theorem consists of showing that
this condition is also sufficient. We rewrite Equation (\ref{equ2})
as
\begin{equation}\label{equ4}
S(t) = \frac{ f'(t) + \nu f(t)}{\beta(t) f(t)},
\end{equation}
then compute $S'(t)$, and then equate with Equation (\ref{equ1}) to
obtain
\begin{equation}\label{equ5}
\frac{d}{dt}\left(\frac{f'(t) + \nu f(t)}{\beta(t) f(t)}\right) = -
\beta(t) \left(\frac{f'(t) + \nu f(t)}{\beta(t) f(t)}\right) f(t).
\end{equation}
Calculating  the derivative and  simplifying the resulting
expression  yields the following Bernoulli differential equation for
$\beta(t)$
\begin{equation}\label{equ6}
 \beta'(t)
- p(t) \beta(t) - f(t) \beta^2(t)=0, \quad \text{where} \quad p(t) =
\frac{{f''(t)}f(t) - {f'(t)}^2}{f(t)({f}'(t)+\nu f(t))}.
\end{equation}
The change of coordinates $x(t)=1/\beta(t)$  transforms this
nonlinear ODE into the linear ODE
\begin{equation}\label{equ7}
{x'(t)} - p(t) x(t) - f(t) =0.
\end{equation}
The method of integrating factors provides the  explicit solution
\begin{equation}\label{equ8} \frac{1}{\beta(t)}= x(t) =  x(0) e^{-P(t)} - e^{-P(t)} \int_0^t e^{P(s)} f(s) ds, \quad
\text{where} \quad P(t) = \int_0^t p(\tau) d\tau.
\end{equation}

A  problem that could arise with this procedure is for the
denominator of $p(t)$ to be zero. A singular solution is prevented
by requiring that the denominator be  always positive, \emph{i.e.},
$f'(t)+\nu f(t)>0$. Having done this, to ensure that $\beta(t)$ is
positive,
 $\beta(0)$ must satisfy
\begin{equation}\label{equ9}
\int_0^T e^{P(s)} f(s) ds<1/\beta(0).
\end{equation}
Mathematically, there are infinitely many choices of $\beta(0)$ and thus
infinitely many transmission functions $\beta(t)$. In this sense the inverse problem is
under-determined.
%This observation
%clearly illustrates a serious danger of overfitting such an epidemic
%transmission model.

\subsection{Recovery algorithm}
We now turn the proof of the theorem into an algorithm to
recover the transmission
rate  $\beta(t)$ from an infection data set. The
algorithm has four steps and requires  two conditions. \vspace{2ex}

\begin{description}
  \item[Step 1.] Smoothly interpolate the infection data with a spline or trigonometric function to generate a
 smooth  $f(t)$. Check condition 1: $\ds f'(t)/f(t) > -\nu$, where $\nu$ is the removal
  rate.
  \item[Step 2.] Compute the function  $\ds p(t)=\frac{{f''(t)}f(t) - {f'(t)}^2}{f(t)({f}'(t)+\nu
  f(t))}$. Condition 1 prevents a zero denominator in $p(t)$.
  \item[Step 3.] Choose $\beta(0)$ and compute the integral $\ds P(t)=\int_0^t p(\tau) d\tau$. Check condition 2: $\ds \beta(0)<1\left/\int_0^T e^{P(s)} f(s) ds\right.$, where $T$
  is the time length of the infection data. Alternatively,
  choose  $\beta(0)$ sufficiently small to satisfy  condition 2.

    \item[Step 4.] Apply the formula $\ds \beta(t)=1\left/\left[e^{-P(t)}/\beta(0) - e^{-P(t)} \int_0^t e^{P(s)} f(s)
  ds\right]\right.$ to compute $\beta(t)$ on the given interval $[0\hspace{1.5ex}T]$.
\end{description}

%Our mathematical derivation of $\beta(t)$ requires a smooth data
%set, since it involves the first and second derivatives of $f(t)$.
%In real life, this is never the case: infection data are
%discrete. Thus, we must first smoothly interpolate the data and then
%apply the explicit formula to the smooth interpolation function.

Condition 1 is equivalent to $d(\ln f(t))/dt>-\nu$, \emph{i.e.}, the
time series of infection data cannot decay too fast at any time.
This is a mild condition that most data sets satisfy. If a
data set does not satisfy this condition, we propose a scaling trick in
Section \ref{Sec:Discussion} to be able to apply the algorithm.

In Section \ref{OtherExtensions}, we present extensions of
the  basic recovery algorithm to several popular extensions of the
SIR model, including the SEIR model with variable vital rates.
Our algorithm can be extended to virtually any such compartment model.

\subsection{Recovering the transmission rate from simulated data}

We first illustrate the recovery algorithm using two simulated data
sets. The functions $f(t)$ and $g(t)$ are the fractions of the
infected population for two characteristic ``types" of infectious
diseases.

The first data set simulates an infectious disease with periodic
outbreaks, as observed in measles (before mass vaccination) and
cholera \citep{ben42,mol95}. The periodic function
$f(t)=10^{-5}[1.4+\cos(1.5 t)]$ represents the continuous infection
data, and Figure 1(a) contains plots of both $f(t)$ (solid) and its
associated transmission rate function $\beta(t)$ (dashed).

The second data set simulates an infectious disease with periodic
outbreaks that decays in time, as observed in influenza
\citep{who01}. The periodic function $g(t)=10^{-5}[1.1 + \sin(t)]
\exp(-0.1 t)$ represents the continuous infection data, and Figure
2(a) contains plots of both $g(t)$ (solid) and its associated
transmission rate function $\beta(t)$ (dashed).

We extract discrete data from functions $f(t)$ and $g(t)$ by
sampling them at equi-spaced intervals (see the small black squares
in Figure 1(a) and Figure 2(a)). To each discrete time series, we
apply two well-known interpolation algorithms (trigonometric
approximation and spline approximation) \citep{kin02,wol09}. Figure
1(b) and Figure 2(b) contain plots of $\beta(t)$ obtained from the
two smooth interpolations together with the recovery algorithm. Both
interpolation schemes yield excellent approximations of $\beta(t)$
in both examples.

Many simulations show that the recovery algorithm is robust with
respect to white noise up to 10\% of the data mean, as well as the
number of sample points.

%Notice that the peaks of $\beta(t)$ in Figure 1 are increasing over
%time. This is a manifestation of the choice of $\beta(0)$ and $\nu$.
%Different values of $\beta(0)$ and $\nu$ may lead to the peaks
%increasing, decreasing, or non-monotone over time. We will see in
%Section \ref{secV} that for historic UK measles and a large range of
%$\beta(0)$, the peaks of $\beta(t)$ are non-decreasing over time.

\subsection{Recovering the transmission rate from UK measles
data}\label{secV}

Previous studies \citep{ear00,kee08} employed the SEIR model with
vital rates to explore the epidemic and endemic behaviors of measles
infections, using the notification data in \citep{opc}. To compare
our new recovery technique with previous measles studies, we extend
our recovery algorithm to the SEIR model with variable vital rates
(see Section \ref{SEIRpre}) and use the same data set. To examine
the robustness of our new results, we post condition the data to
account for underreporting and reapply the extended recovery
algorithm.

We use the measles parameter values from \citet{and92,opc}:
$\nu=52$/year$=52/12$/month (where $1/\nu$ is the removal period),
$a=52$/year$=52/12$/month (where $1/a$ is the latent period), and
$\delta=1/70$/year$=1/70/12$/month (death rate, equivalent to the
life span of 70 years).

Public databases, such as the International Infectious Disease Data
Archive \citep{iidda} and Bolker's measles data archive
\citep{bolker}, contain the weekly numbers of measles notifications
from $1948-1966$ and the quarterly reported historical UK births
from $1948-1956$. During $1948 -1956$ the births show large annual
variations (see Figure 3(c)) with a strong $1$/year frequency
component (see Figure 3(d)). Since some years these variations approach
$20\%$,  we  include actual births in our
model. Since neither database contains the UK birth rates from
$1957-1966$, this requires us to restrict our study to the period $1948-1956$.

Although disease notification data (the number of reported new
infections during a given period) is different from prevalence data
(the total number of infections during the period), assuming all
infections are reported, they should be close if the reporting
period is longer than the mean generation time of infection.
Previous measles modelers have used notification data as a surrogate
for the number of infected individuals \citep{yor73,ear00}. To be
able to compare our results with those of pervious authors, we first
apply our recovery algorithm to the same notification data and then
check the robustness of our algorithm by applying it to estimated
prevalence data. We thank David Earn for clarifying these issues for
us.

Since the birth data is provided only quarterly and the
notifications weekly, we smoothly interpolate the birth data and
aggregate the notification data into one month intervals.
%The aggregation provides a type of low pass filter.
To aggregate weekly infection
data into monthly data, we simply sum the  weekly data as previous
studies \citep{yor73,ear00}. For a week across two months, this
weekly infection number is separated to be two parts. For instance,
if one week has three days in May and four days in June, then we
multiply the notification data of this week by $3/7$ and incorporate
it into May data, and we multiply the notification data of this week
by $4/7$ and incorporate it into June data.

In Figure 4(a)(c)(e)(g), we plot the transmission rates $\beta(t)$
recovered from our algorithm for four different January initial
values chosen to represent a wide range of $\beta(0)$. The recovered
$\beta(t)$ in panels (e)(g) are stationary, slowly increasing peaks
in panel (c), and fast increasing peaks in panel (a). Note that
annual minima occur in July-August during the summer school vacation
period and that annual maxima occur in January or September during
the first month after the winter and summer school vacations.
%The
%basic reproduction ratio, $R_0$, can be estimated by $\beta/\nu$.
%Notice that $\beta$ in our scaled model is equivalent to $\beta N$
%in the unscaled model, where $N$ is the total population. For the
%stationary case in Figure 4(e), $R_0$ varies from $1.3$ to $4.6$.

In Figure 4(b)(d)(f)(h), we plot the moduli of Fourier transform of
all recovered $\beta(t)$ and observe that there are two competing
dominant spectral peaks. These two dominant peaks have $1$ and
$1/3$-year periods. The three per year (\emph{i.e.}, $1/3$-year
period) peak of $\beta(t)$ is due to the three major school terms,
which are separated by the Christmas, Easter, and Summer breaks (see
Figure 6). For stationary $\beta(t)$ (see panels (e)(g)), the one
per year spectral peak is dominant (see panels (f)(h)), and the one
half per year ($2$-year period) spectral peak is
comparable to the three per year spectral peak (see panel (h)).
We conclude that for a huge range of $\beta(0)$, the
transmission rate always possesses both strong $1$ and $1/3$-year cycles.

To test the robustness of our spectral peaks, we incorporate the
standard correction factor of $92.3\%$ to account for the
underreporting bias in the UK measles data (with estimated mean
reporting rate $52\%$, note that $92.3\%$ is computed from
$1/0.52-1$) \citep{cla85,bjo02,xia04}. Since both the removal stage
in the SEIR model and the data notification period are one week, and
the $92.3\%$ correction factor essentially doubles the number of
cases, this corrected number of cases will account for the
non-notified cases during this removal stage. Hence, the corrected
weekly notification data should provide a good approximation for
total weekly infections. This precise methodology was used in
\citep{bol93,bjo02}.

In   Figure 5. we plot the recovered $\beta(t)$ with the $92.3\%$ correction factor
and for the large range of $\beta(0)$. All the recovered
$\beta(t)$ have identical spectral peaks as those in Figure 4. Thus
our observation of the two spectral peaks with frequencies $1$/year
and $3$/year seems robust. Again, we observe that the scale of the infection data regulates
the scale of $\beta(t)$ but does not affect spectral peaks of
Fourier transform of $\beta(t)$.

Most previous measles models represent the time-varying transmission rate
using a school term-time forcing function (such as sinusoidal or Haar
function), and fitted parameters using the method of least squares,
without providing variances for their estimates \citep{bol93}.
Little empirical evidence currently supports their assumed functions. The
school mixing assumption behind these simple transmission functions
ignores many other seasonal factors such as environmental changes
and immune system changes \citep{fuj98,pat00}.  However, our
 observations support  this common belief that measles transmission is
predominantly driven by school contacts.

%
%Previous authors xxxxx\label{eqn:xxx}   The transmission rate
%has a huge variation as tested numerically and assumed in previous
%studies \citep{ear00,bjo02,ell98,kee08}. Even the ballpark range of
%the transmission rate function is highly uncertain
%\citep{lon73,yor73}, in part since it depends on guessing the size
%of susceptible population.

\section{Discussion}\label{Sec:Discussion}

We present a new algorithm to compute the time-dependent transmission rate from prevalence data, which makes no assumptions about the number of susceptibles or vital rates. We do have to  estimate $\beta(0)$, which can be a formidable challenge. By manipulating our derivation, Hadeler recently derived an even simpler inversion
formula for $\beta(t)$ that requires knowledge of $S(0)$ instead of
$\beta(0)$ \citep{had10}.

Our algorithm
can be viewed as a continuous version of the well-known discrete
method for estimating the time-varying transmission rate. The
inverse method in this paper can be applied to derive recovery
algorithms for a large spectrum of epidemiological models (see
Section IV B). In this sense our new method provides a more general
method than the discrete method since it can account for factors
such as the transmission mode and the immunity memory period which
can have a significant effect on   transmission rate.

We illustrate the recovery algorithm for the SEIR model with
variable vital rates using UK measles data from 1948 to 1956.
Fourier transform of our recovered transmission rate function shows
dominant spectral peaks at frequencies $1$ and $3$ per year. The $3$
per year frequency arises from three major school breaks.  Our
 observations support  the common belief that measles transmission is
predominantly driven by school contacts, but we also find indications of a
two year cycle.

%In
%addition, our recovery algorithm provides a new method to estimate
%$\beta(0)$ which does not involve guestimating $S(0)$.

Our algorithm has some limitations to its applicability. First, the
proportion of infected individuals, $f(t)$, can not decrease too
fast over the full time interval of interest. In general, one can
add a sufficiently large constant to $f(t)$ to ensure this, but this
will change the range of applicable $\beta(0)$, and applicability
needs to be checked. Second, one must assume that the proportion (or
number) of notifications is always strictly positive. In practice
this restriction can be overcome by replacing zero values in the
time series with a very small positive value. Third, for a chosen
$\beta(0)$, the algorithm can only apply to a finite length of
infection data. Finally, one either needs to know the value of the
transmission rate at some fixed time, or verify that the desired
properties of $\beta(t)$ hold for all $\beta(0)$ in the range where
the estimated $\beta(t)$ is stationary.

The algorithm should apply to the vast majority of infection data
sets, and a consequence is that one can nearly always construct a
time-dependent transmission rate $\beta(t)$ such that SIR model will
fit the data perfectly. This illustrates a potential danger of
overfitting an epidemic model with time-dependent transmission rate.

Recently, likelihood-based methods for estimating the time-varying
transmission function have been developed by \citet{cau08,he09}
using stochastic transmission models. Our algorithm can also be
extended to incorporate stochastic effects by using stochastic
differential equations.

\section{Materials and Methods}

\subsection{Extensions of the basic model}\label{OtherExtensions}

Analogous results and inversion formulae hold for all standard variations of the standard
SIR model and their combinations. The proofs are very similar to the
proof of Theorem  (\ref{theory2}). Here, we only
present the full algorithm for the SEIR model with vital rates, since we
apply this  algorithm to UK measles data.

\subsubsection{SIR model with vital rates} \vspace{-.5in}
\begin{eqnarray}\label{equ20}
  S'(t) &=&  \delta-\beta(t) S(t)I(t)-\delta S(t), \\\label{equ21}
  I'(t) &=& \beta(t) S(t)I(t)-\nu I(t)-\delta I(t), \\\label{equ22}
  R'(t) &=& \nu I(t)-\delta R(t).
\end{eqnarray}
The necessary and sufficient condition for recovering $\beta(t)$
given $\nu$ and $\delta$ is $ f'(t)/f(t) > -(\nu + \delta)$.

\subsubsection{SIR model with  waning immunity}\vspace{-.5in}

\begin{eqnarray}\label{equ14}
  S'(t) &=& m R(t)-\beta(t) S(t)I(t), \\\label{equ15}
  I'(t) &=& \beta(t) S(t)I(t)-\nu I(t), \\\label{equ16}
  R'(t) &=& \nu I(t)- m R(t),
\end{eqnarray}
where $1/m$ is the memory period of immunity. The necessary and
sufficient condition for recovering $\beta(t)$ given $\nu$ is
$f'(t)/f(t) > -\nu$.

\subsubsection{SIR model with time-dependent indirect transmission
rate (\citet{ric09})}\vspace{-.5in}
\begin{eqnarray}\label{equ10}
  S'(t) &=& -\omega(t) S(t), \\\label{equ11}
  I'(t) &=& \omega(t) S(t)-\nu I(t), \\\label{equ12}
  R'(t) &=& \nu I(t),
\end{eqnarray}
where $\omega(t)$ is the time-dependent indirect transmission rate.
The necessary and sufficient condition for recovering $\beta(t)$
given $\nu$ is $f'(t)/f(t) > -\nu$.

\subsubsection{SEIR model}\label{SEIRabs}\vspace{-.5in}
\begin{eqnarray}\label{eqn:dotSv}
  S'(t)  &=& -\beta(t) S(t)I(t), \\\label{eqn:dotEv}
  E'(t)  &=& \beta(t) S(t)I(t) - \alpha E(t), \\\label{eqn:dotIv}
  I'(t)  &=& \alpha E(t) - \nu I(t), \\\label{eqn:dotRv}
  R'(t)  &=& \nu I(t),
\end{eqnarray}
where $1/\alpha$ is the latent period for the disease. By simple
calculations, we can show that the necessary and sufficient
condition for recovering $\beta(t)$ from infection data is
$f'(t)/f(t) > -\nu$.

\subsubsection{SEIR model with  vital
rates}\label{SEIRpre}\vspace{-.5in}
\begin{eqnarray}\label{equ30}
 S'(t) &=& \delta-\beta(t) S(t)I(t)-\delta S(t), \\\label{equ31}
 E'(t) &=& \beta(t) S(t)I(t)-a E(t)-\delta E(t), \\\label{equ32}
 I'(t) &=& a E(t)-\nu I(t)-\delta I(t), \\\label{equ33}
 R'(t) &=& \nu I(t)-\delta R(t).
\end{eqnarray}
The necessary and sufficient conditions for recovering $\beta(t)$
from infection data are
\begin{equation}\label{equ34}
{f'(t)}+(\nu+\delta)f(t)>0 \quad \text{and} \quad
{f''(t)}+(\nu+2\delta+a){f'(t)}+(\delta+a)(\nu+\delta)f(t)>0.
\end{equation}
In this case, $\beta(t)$ satisfies the Bernoulli equation
\begin{equation}\label{equ24}
{\beta'}+p(t)\beta+q(t)\beta^2=0,
\end{equation}
where
\begin{eqnarray*}
p(t) &=&
\frac{-a{f'''(t)}f(t)-a(\nu+2\delta+a){f''(t)}f(t)-a(\delta+a)(\nu+\delta){f'(t)}f(t)+a{f''(t)}{f'(t)}+a(\nu+2\delta+a){f'(t)}^2
}{af(t)[{f''(t)}+(\nu+2\delta+a){f'(t)}+(\delta+a)(\nu+\delta)f(t)]} \\
 &+& \frac{a(\delta+a)(\nu+\delta){f'(t)}f(t)-\delta a{f''(t)}f(t)-\delta a(\nu+2\delta+a){f'(t)}f(t)-\delta a(\delta+a)(\nu+\delta)f^2(t)}{af(t)[{f''(t)}+(\nu+2\delta+a){f'(t)}+(\delta+a)(\nu+\delta)f(t)]},
\end{eqnarray*}
and
\begin{eqnarray*}
q(t)  &=&  \frac{\delta a^2 f^2(t)-a
{f''(t)}f^2(t)-a(\nu+2\delta+a){f'(t)}f^2(t)-a(\delta+a)(\nu+\delta)f^3(t)}{af(t)[{f''(t)}
+(\nu+2\delta+a){f'(t)}+(\delta+a)(\nu+\delta)f(t)]}.
\end{eqnarray*}

\vspace{2ex}

The modified recovery algorithm has five steps together with three
conditions.

\begin{description}
  \item[Step 1.] Smoothly interpolate the infection data to generate a
  smooth  function $f(t)$ that has at least a continuous second
  derivative. Check condition 1: ${f'(t)}+(\nu+\delta)f(t)>0$; and check condition 2:
  ${f''(t)}+(\nu+2\delta+a){f'(t)}+(\delta+a)(\nu+\delta)f(t)>0$.
  \item[Step 2.] Compute the function $p(t) = \frac{-a{f'''(t)}f(t)-a(\nu+2\delta+a){f''(t)}f(t)-a(\delta+a)(\nu+\delta){f'(t)}f(t)+a{f''(t)}{f'(t)}+a(\nu+2\delta+a){f'(t)}^2}
  {af(t)[{f''(t)}+(\nu+2\delta+a){f'(t)}+(\delta+a)(\nu+\delta)f(t)]}$
 $+\frac{a(\delta+a)(\nu+\delta){f'(t)}f(t)-\delta a{f''(t)}f(t)-\delta a(\nu+2\delta+a){f'(t)}f(t)-\delta
 a(\delta+a)(\nu+\delta)f^2(t)}{af(t)[{f''(t)}+(\nu+2\delta+a){f'(t)}+(\delta+a)(\nu+\delta)f(t)]}.$
  \item[Step 3.] Choose $\beta(0)$ and compute the integral $\ds P(t)=\int_0^t p(\tau) d\tau$.
  Check condition 3: $$\ds \frac{1}{\beta(0)}+\int_0^T e^{-P(s)} q(s)
  ds>0.$$
  \item[Step 4.] Compute the function $q(t) = \frac{\delta a^2 f^2(t)-a
{f''(t)}f^2(t)-a(\nu+2\delta+a){f'(t)}f^2(t)-a(\delta+a)(\nu+\delta)f^3(t)}{af(t)[{f''(t)}
+(\nu+2\delta+a){f'(t)}+(\delta+a)(\nu+\delta)f(t)]}$.
  \item[Step 5.] Apply the formula $\ds \beta(t)=1\left/\left[e^{P(t)}/\beta(0) + e^{P(t)} \int_0^t e^{-P(s)} q(s)
  ds\right]\right.$ to compute $\beta(t)$ on the given interval $[0\hspace{1.5ex}T]$.
\end{description}

With variable birth rate $\eta(t)$ and constant death rate $\delta$,
then the SEIR model becomes
\begin{eqnarray}\label{equ101}
 S'(t) &=& \eta(t)-\beta(t) S(t)I(t)-\delta S(t), \\\label{equ31}
 E'(t) &=& \beta(t) S(t)I(t)-a E(t)-\delta E(t), \\\label{equ32}
 I'(t) &=& a E(t)-\nu I(t)-\delta I(t), \\\label{equ33}
 R'(t) &=& \nu I(t)-\delta R(t).
\end{eqnarray}
In this case, the formula in Step 4 should be $$q(t) = \frac{\eta(t)
a^2 f^2(t)-a
{f''(t)}f^2(t)-a(\nu+2\delta+a){f'(t)}f^2(t)-a(\delta+a)(\nu+\delta)f^3(t)}{af(t)[{f''(t)}
+(\nu+2\delta+a){f'(t)}+(\delta+a)(\nu+\delta)f(t)]}.$$ All other
steps in the algorithm remain the same.

\section*{Acknowledgments}

\noindent The UK measles data and birth data was obtained from
\citep{iidda} and \citep{bolker}. The authors very much appreciate
the efforts of David Earn and Ben Bolker for maintaining these
public databases of infectious disease data. Thank David Earn for
helpful discussion. The authors also thank an anonymous referee for
identifying the source of the three times per year frequency
component in our approximation of the transmission rate for measles.

\newpage

Figure 1. (a) We extract $21$ equally spaced  data points from the
periodic function $f(t)=10^{-5}[1.4+\cos(1.5 t)]$; the dashed curve
is $\beta(t)$ recovered from $f(t)$ using (\ref{equ8}). (b) These
transmission functions are estimated using spline and trigonometric
interpolations on the $21$ data points. \vspace{3ex}

Figure 2. (a) We extract $21$ equally spaced data points from the
oscillatory decaying function $g(t)=10^{-5}[1.1 + \sin(t)] \exp(-0.1
t)$; the dashed curve is $\beta(t)$ recovered from $g(t)$ using
(\ref{equ8}). (b) These transmission functions are estimated using
spline and trigonometric interpolations on the $21$ data points.
\vspace{3ex}

Figure 3. (a) Aggregated monthly measles data from England and Wales
in $1948-1956$. (b) Fourier transform of filtered and smoothly
interpolated aggregated monthly data showing the dominant frequency
components (normalized modulus). Note: we filter and remove the
artificial peak at zero frequency in Fourier transform. (c) UK birth
rates during $1948-1956$. (d) Fourier transform of smoothly
interpolated UK birth data showing the dominant frequency component
(normalized modulus). \vspace{3ex}

Figure 4. The transmission rate function $\beta(t)$ recovered from
our extended algorithm with historic birth rates. (a) The recovered
$\beta(t)$ with $\beta(0)=270$: fast increasing peaks. (b) Fourier
transform of filtered $\beta(t)$ showing the dominant frequency
component, $3$ per year. (c) The recovered $\beta(t)$ with
$\beta(0)=230$: slowly increasing peaks. (d) Fourier transform of
filtered $\beta(t)$ showing two comparable dominant frequencies
components, $1$ and $3$ per year. (e) The recovered $\beta(t)$ with
$\beta(0)=140$: stationary peaks. (f) Fourier transform of filtered
$\beta(t)$ showing the dominant frequency component, $1$ per year.
(g) The recovered $\beta(t)$ with $\beta(0)=80$: stationary peaks.
(h) Fourier transform of filtered $\beta(t)$ showing the dominant
frequency component, $1$ per year; $1/2$ per year peak is large and
comparable to $3$ per year peak. \vspace{3ex}

Figure 5. We test our estimations of $\beta(t)$ with data
correction. The moduli of Fourier transform of $\beta(t)$ with the
$92.3\%$ correction factor have identical spectral peaks as those
without data correction in Figure 4. (a) The recovered $\beta(t)$
with $\beta(0)=120$. (c) The recovered $\beta(t)$ with
$\beta(0)=100$. (e) The recovered $\beta(t)$ with $\beta(0)=60$. (g)
The recovered $\beta(t)$ with $\beta(0)=30$. Panels (b)(d)(f)(h)
plot moduli of Fourier transform of corresponding $\beta(t)$ in
(a)(c)(e)(g), respectively. \vspace{3ex}

Figure 6. The transmission rate $\beta(t)$ is generated from Haar
function with major school holidays via aggregation. The modulus of
Fourier transform of $\beta(t)$ generated from the widely used
periodic Haar function shows the dominant three times per year
frequency.

\newpage

\begin{figure}[h]
\begin{center}$
\begin{array}{cc}
\subfigure[] {\includegraphics[width=2.8in]{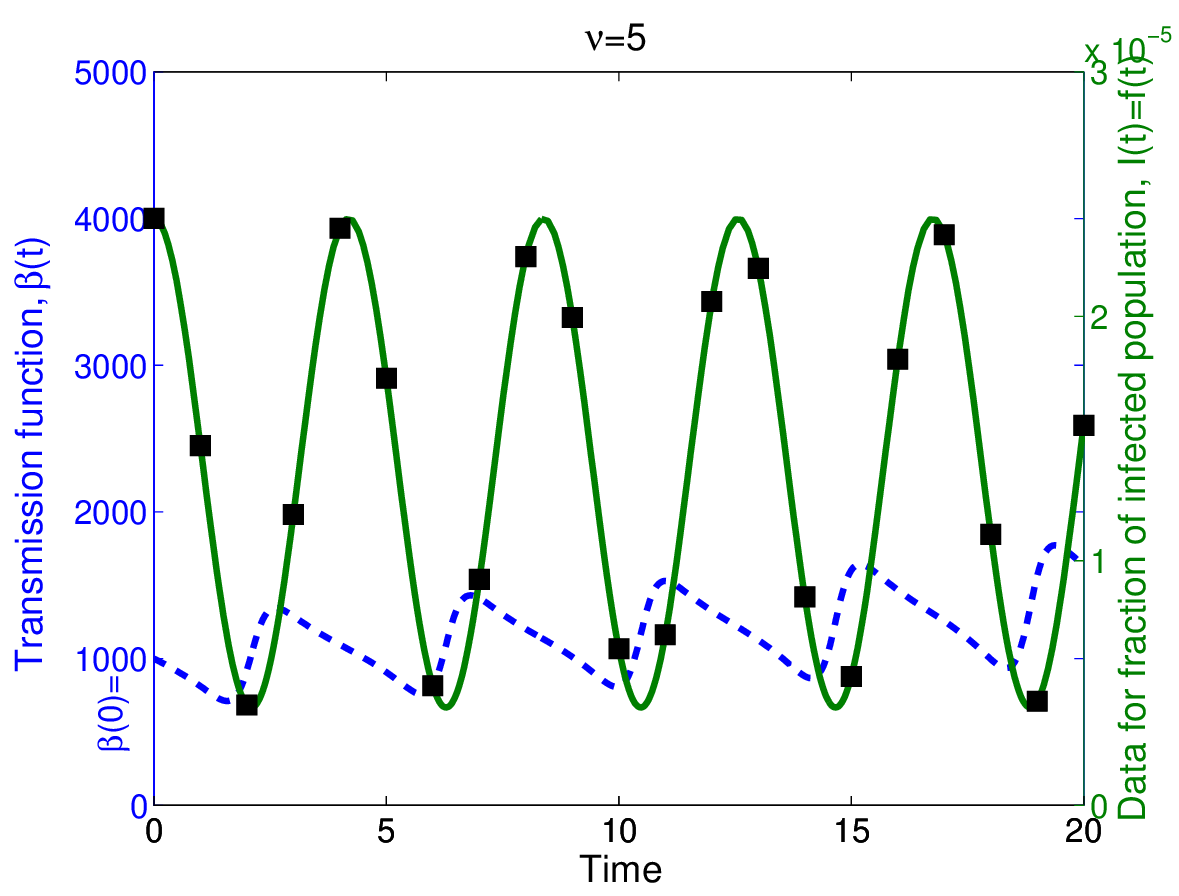}}\\
\subfigure[]
{\includegraphics[width=3.2in]{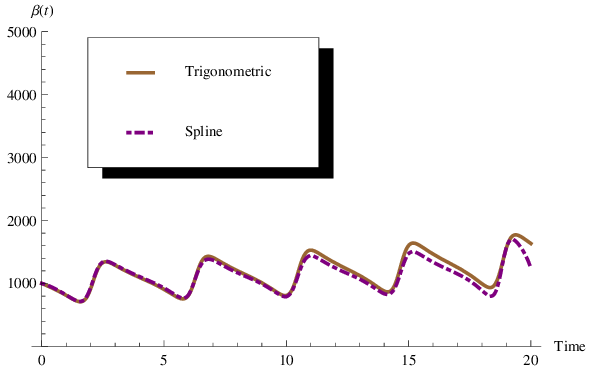}}
\end{array}$
\end{center}
\caption{}\label{fig:example1}
\end{figure}

\newpage

\begin{figure}[h]
\begin{center}$
\begin{array}{cc}
\subfigure[] {\includegraphics[width=2.8in]{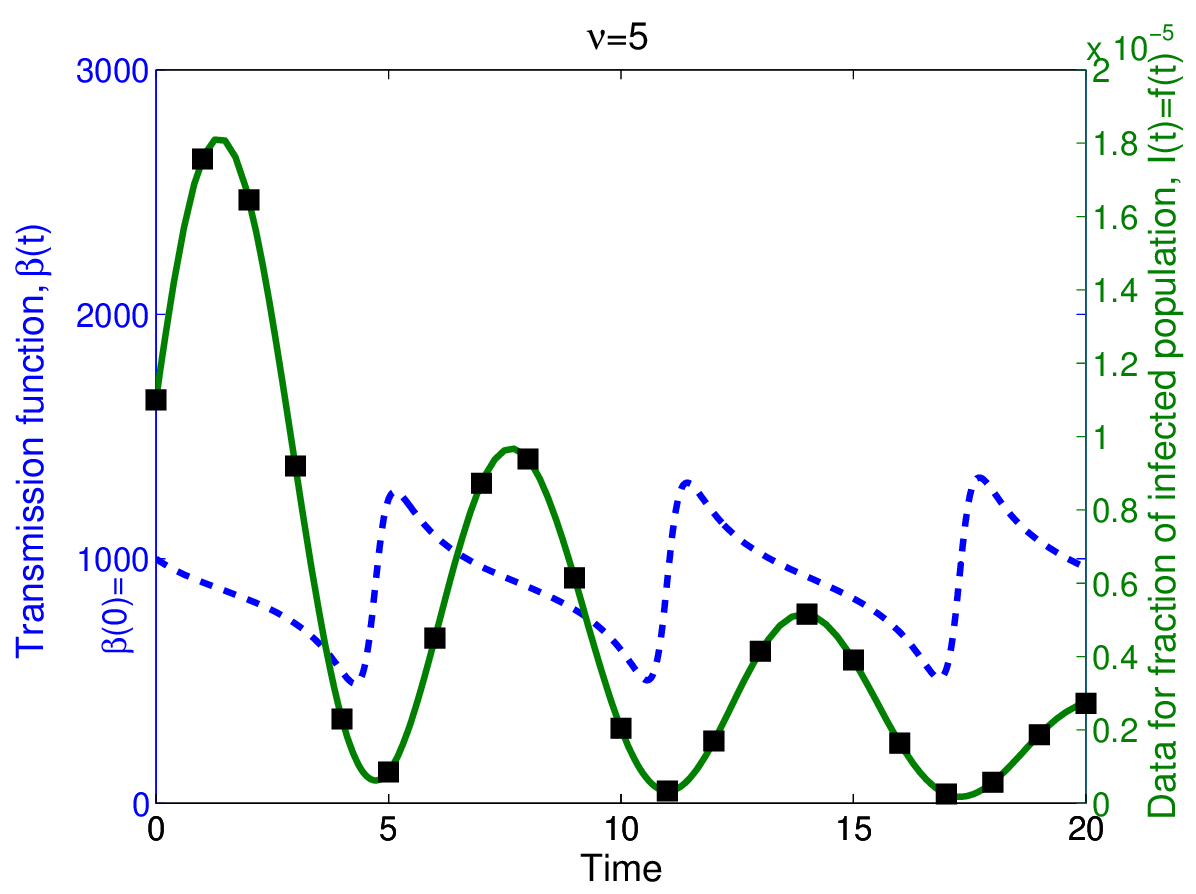}}\\
\subfigure[]
{\includegraphics[width=3.2in]{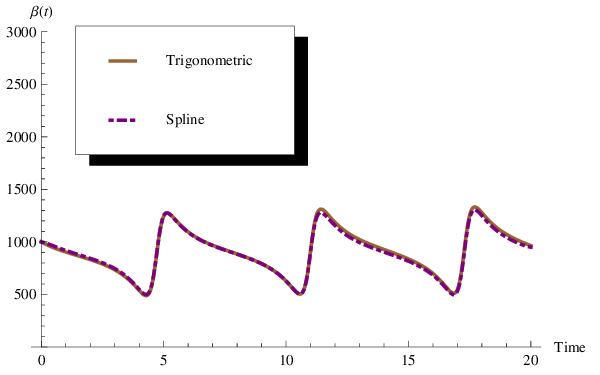}}
\end{array}$
\end{center}
\caption{}\label{fig:example2}
\end{figure}

\newpage

\begin{figure}[h]
\begin{center}$
\begin{array}{cccc}
\subfigure[] {\includegraphics[width=2.1in]{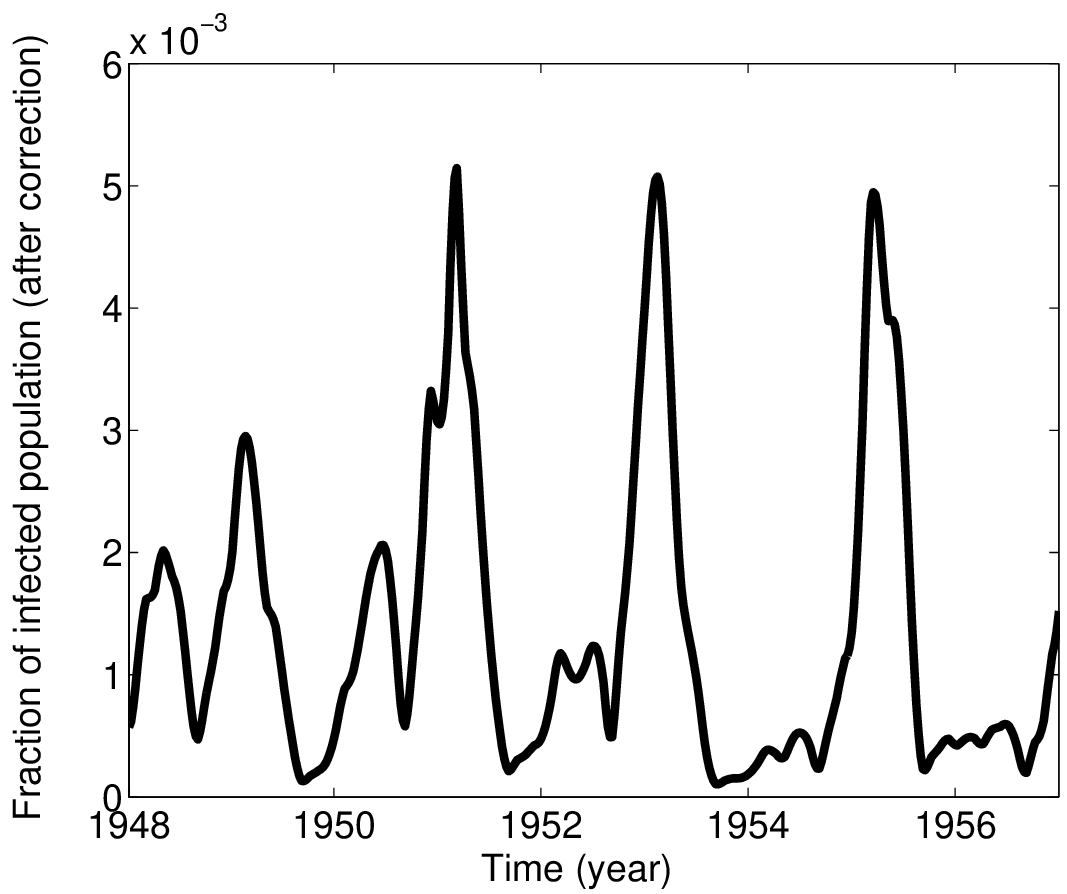}}
\subfigure[] {\includegraphics[width=2.2in]{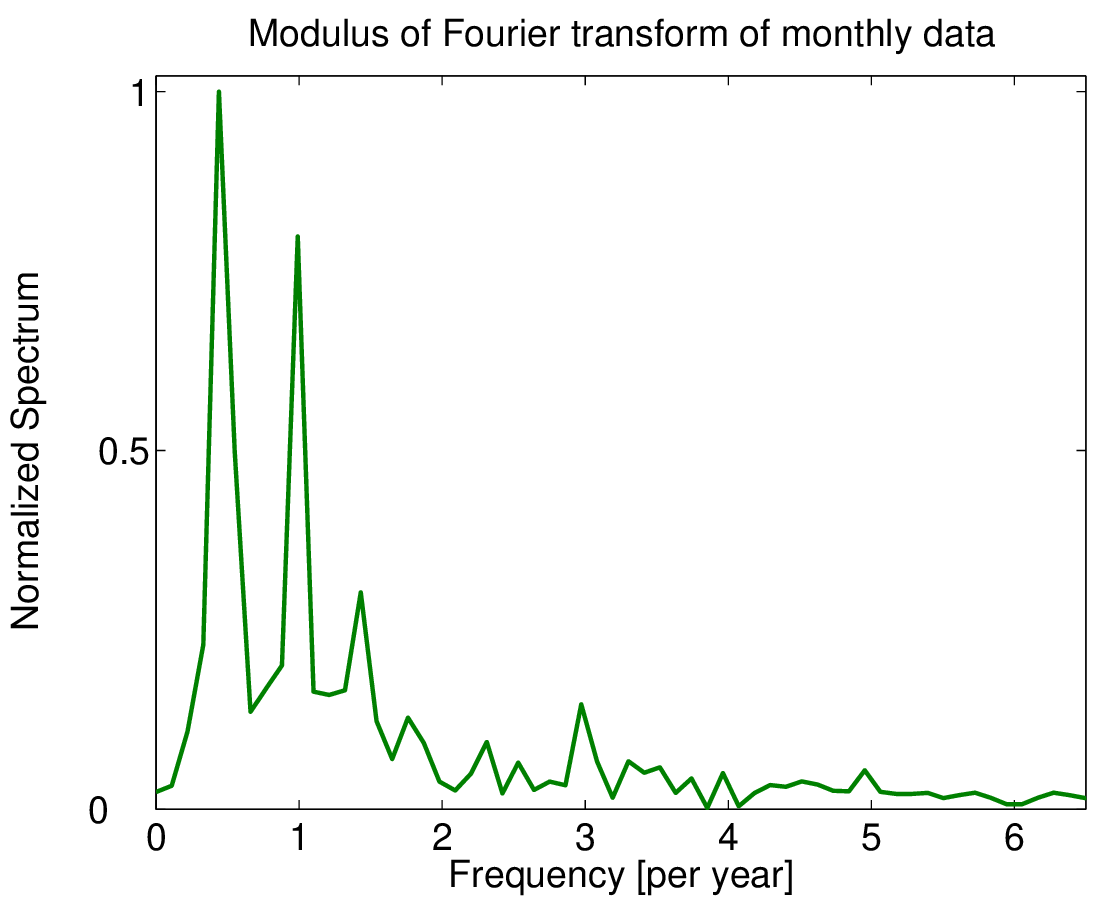}}\\
\subfigure[] {\includegraphics[width=2.3in]{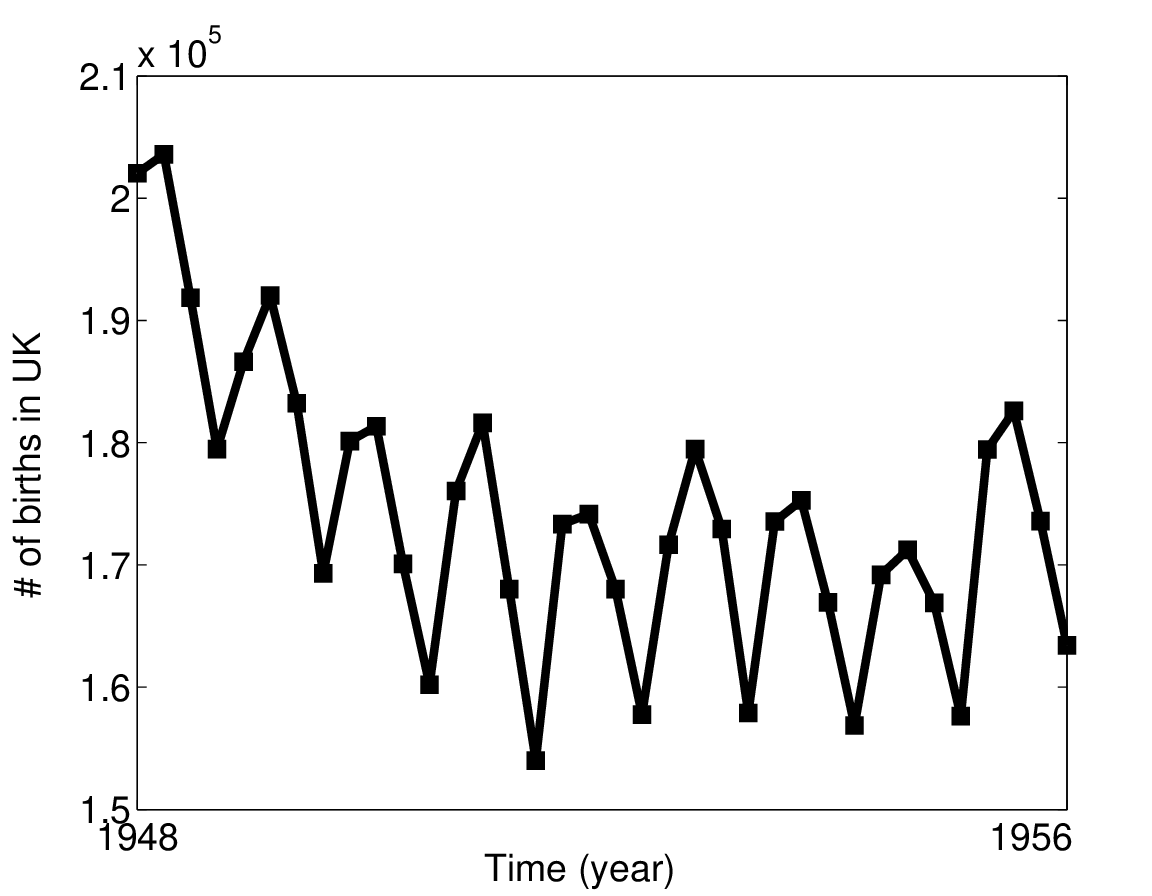}}
\subfigure[]
{\includegraphics[width=2.1in]{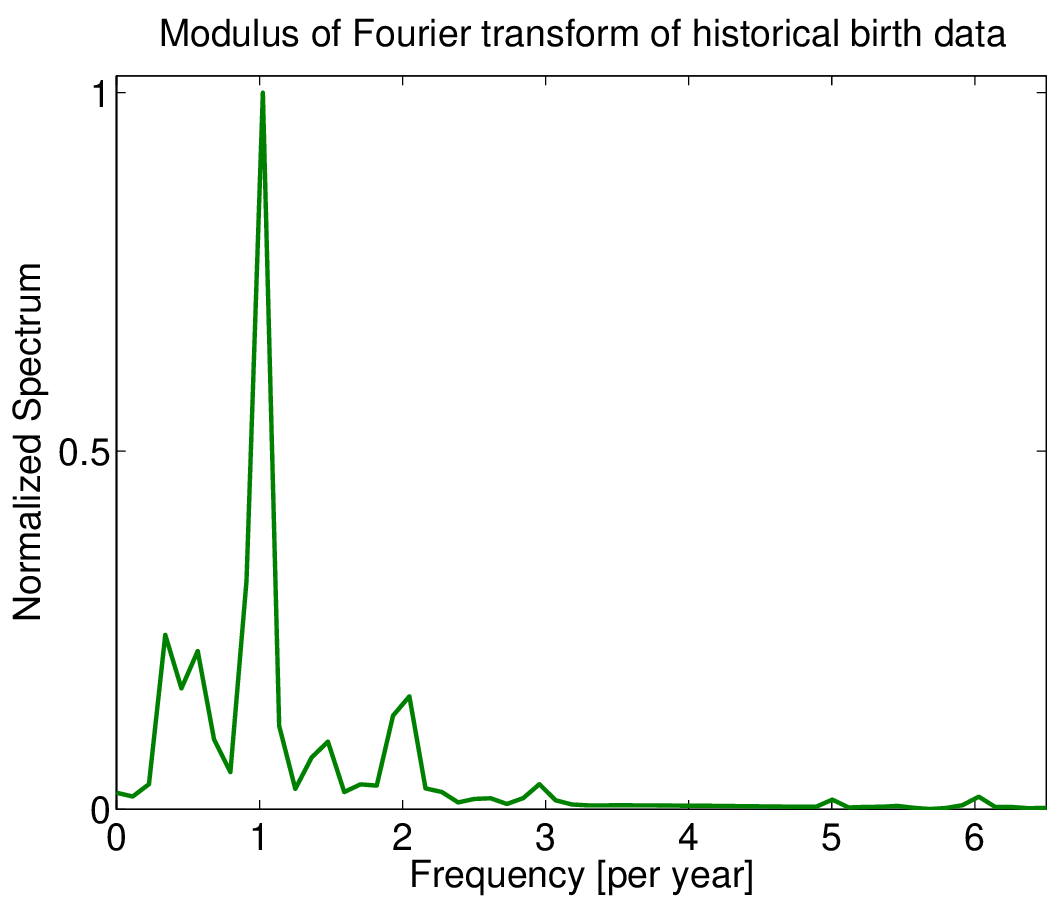}}
\end{array}$
\end{center}
\caption{}\label{fig:spectrum}
\end{figure}

\begin{figure}[h]
\begin{center}$
\begin{array}{ccc}
\subfigure[] {\includegraphics[width=2.5in]{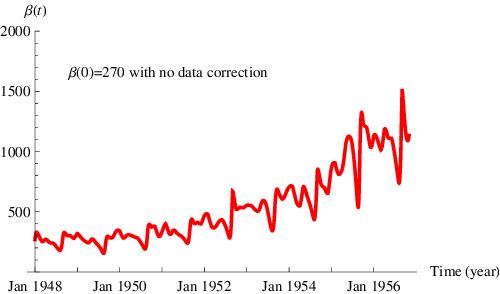}}
\subfigure[] {\includegraphics[width=2in]{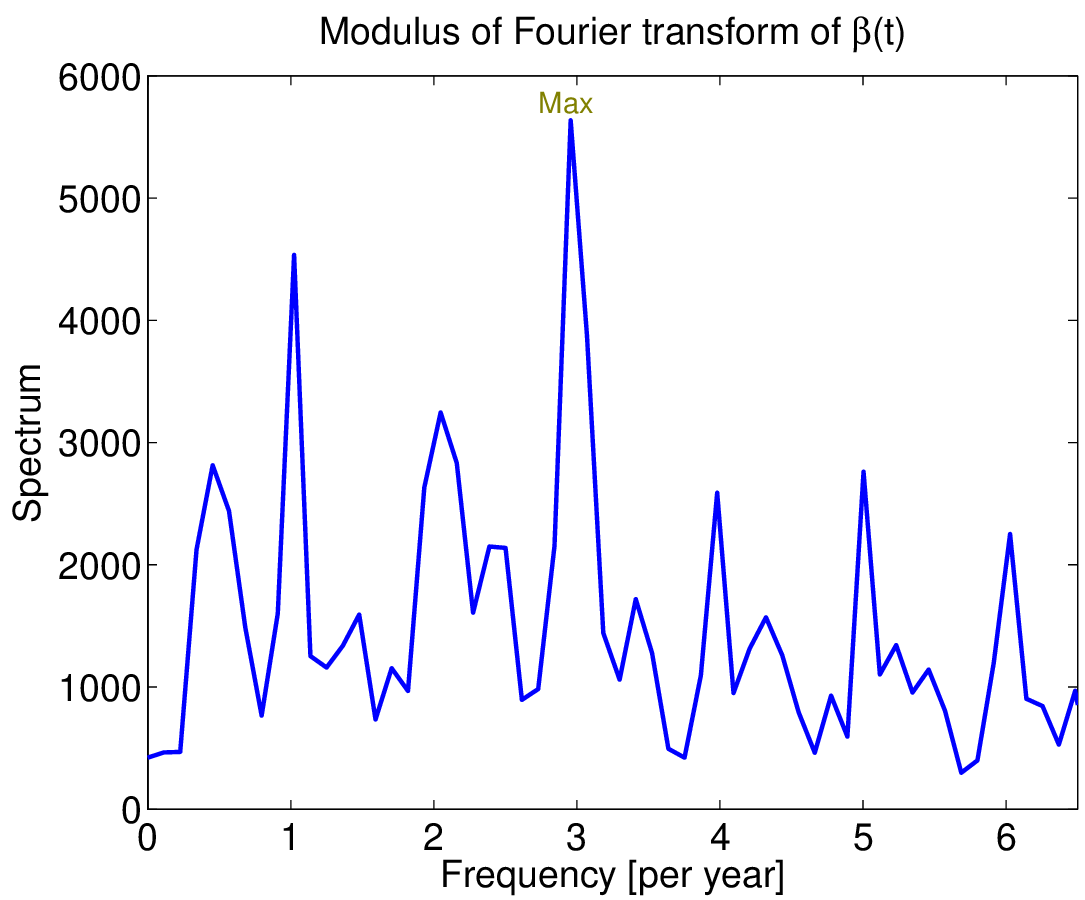}}\\
\subfigure[] {\includegraphics[width=2.5in]{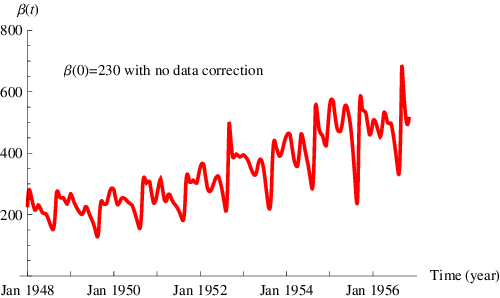}}
\subfigure[] {\includegraphics[width=2in]{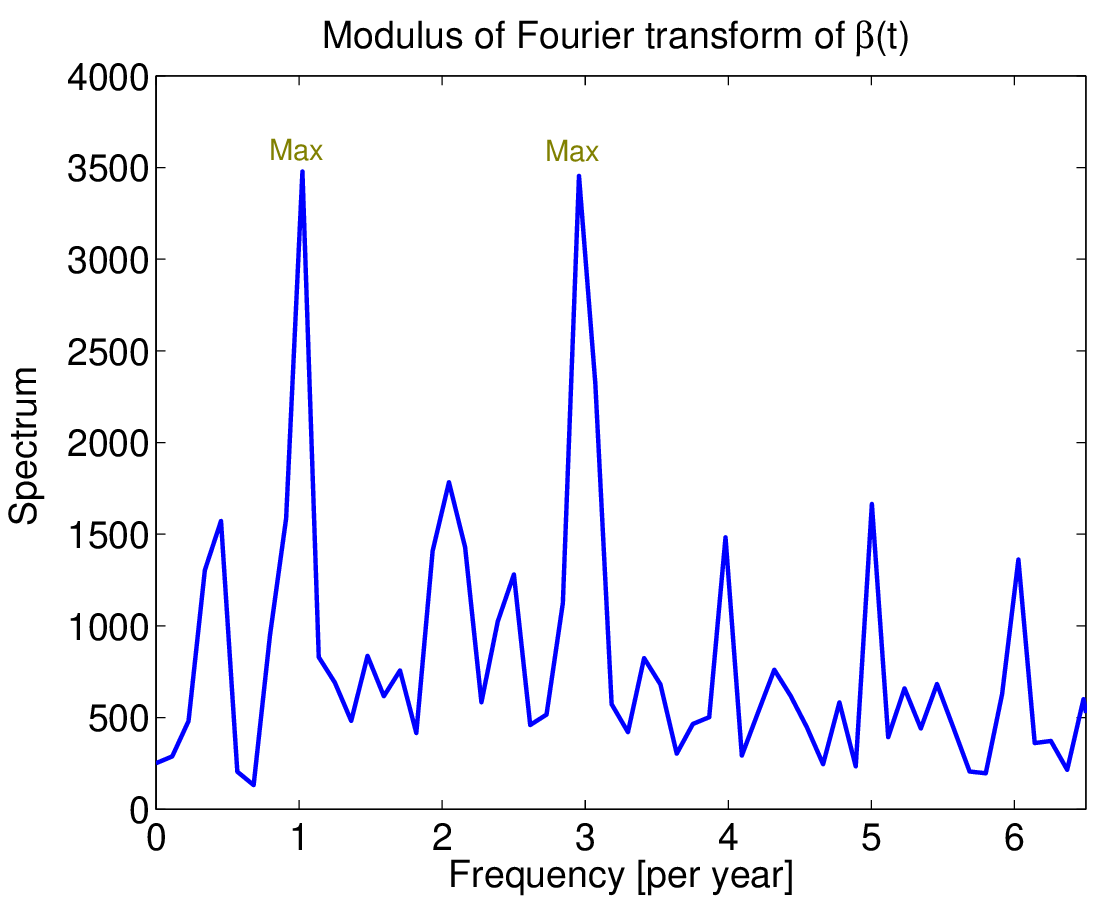}}\\
\subfigure[] {\includegraphics[width=2.5in]{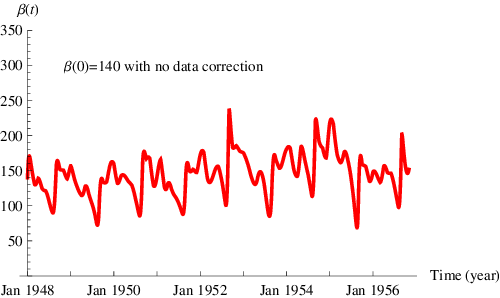}}
\subfigure[]
{\includegraphics[width=2in]{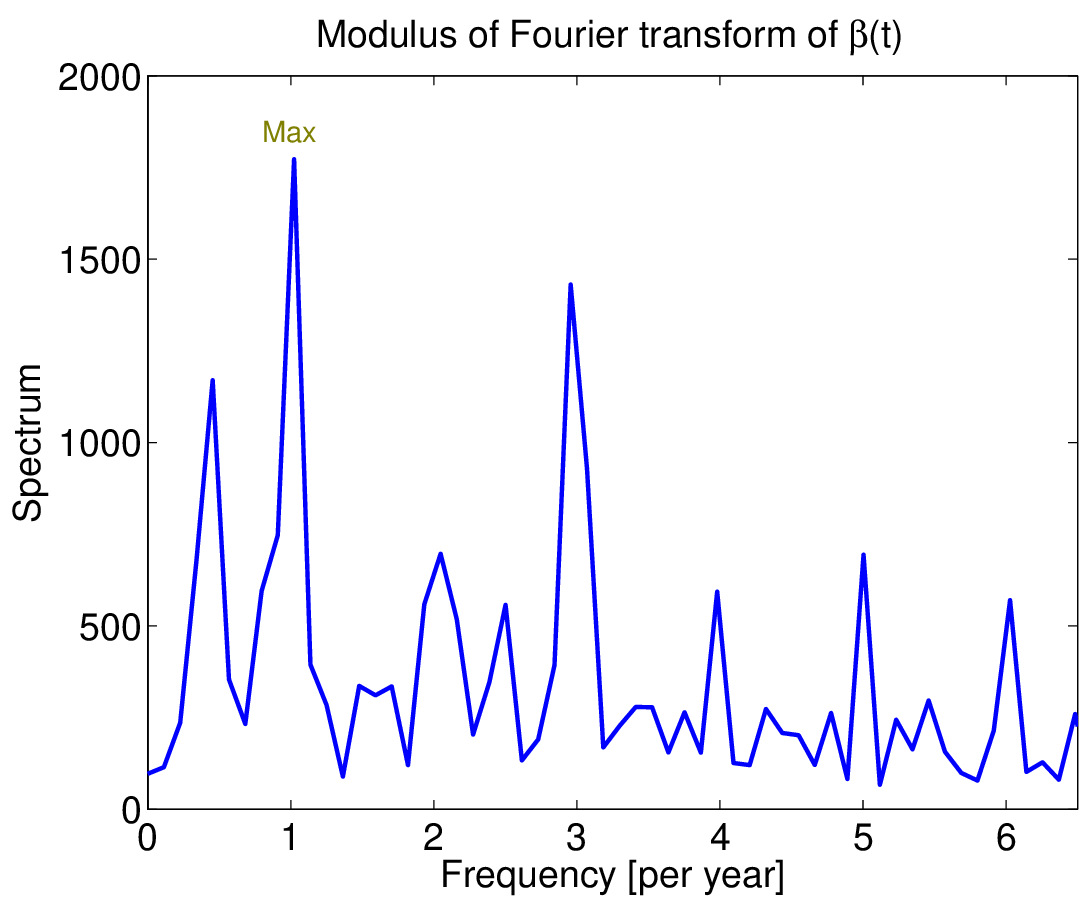}}\\
\subfigure[] {\includegraphics[width=2.5in]{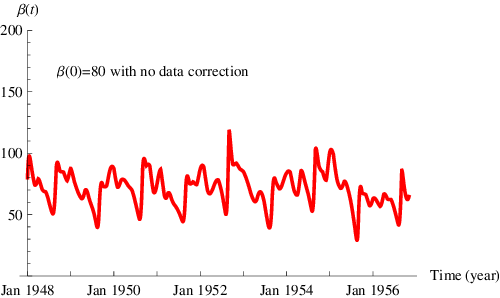}}
\subfigure[]
{\includegraphics[width=2in]{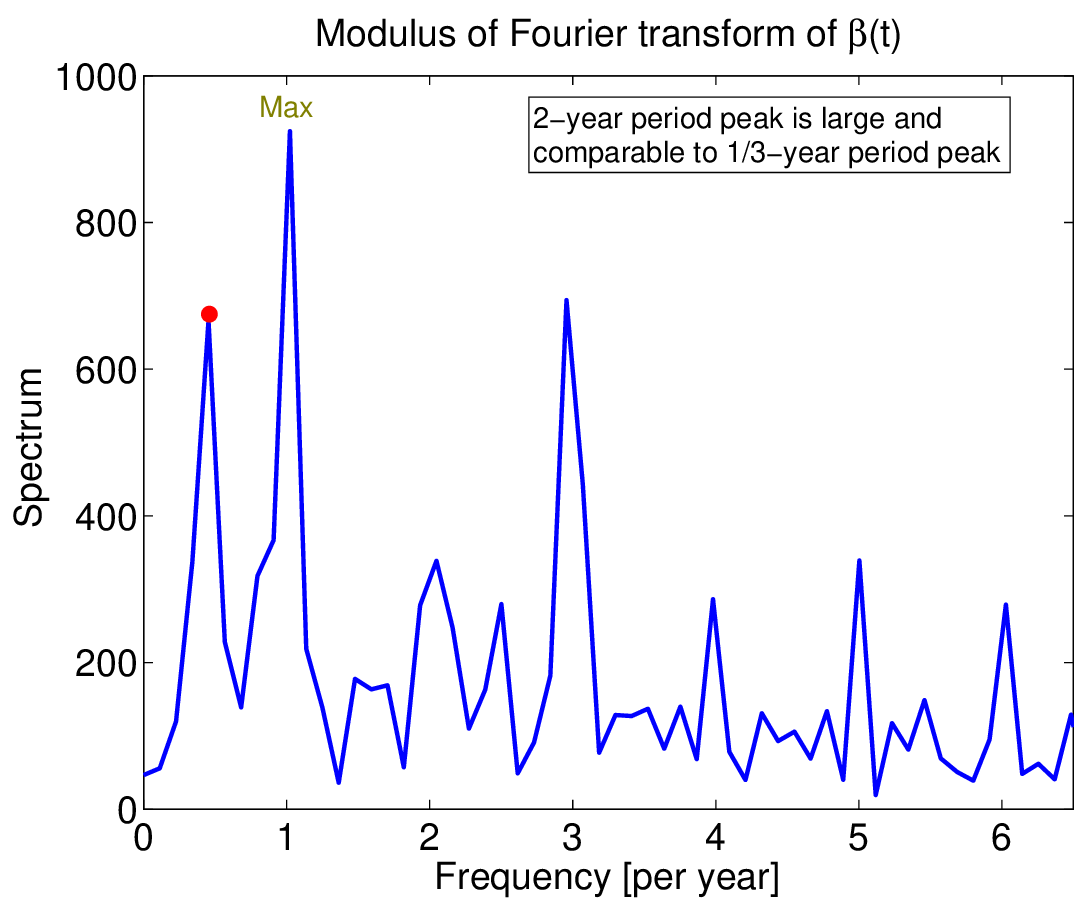}}
\end{array}$
\end{center}
\caption{}\label{fig:NoCorrectionTest}
\end{figure}

\begin{figure}[h]
\begin{center}$
\begin{array}{cccccc}
\subfigure[] {\includegraphics[width=2.5in]{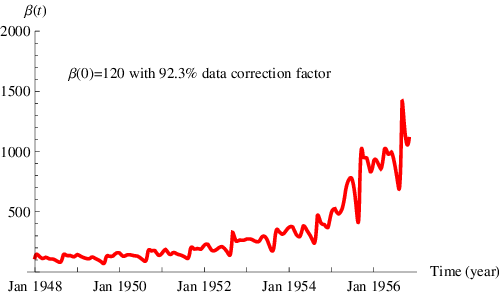}}
\subfigure[] {\includegraphics[width=2in]{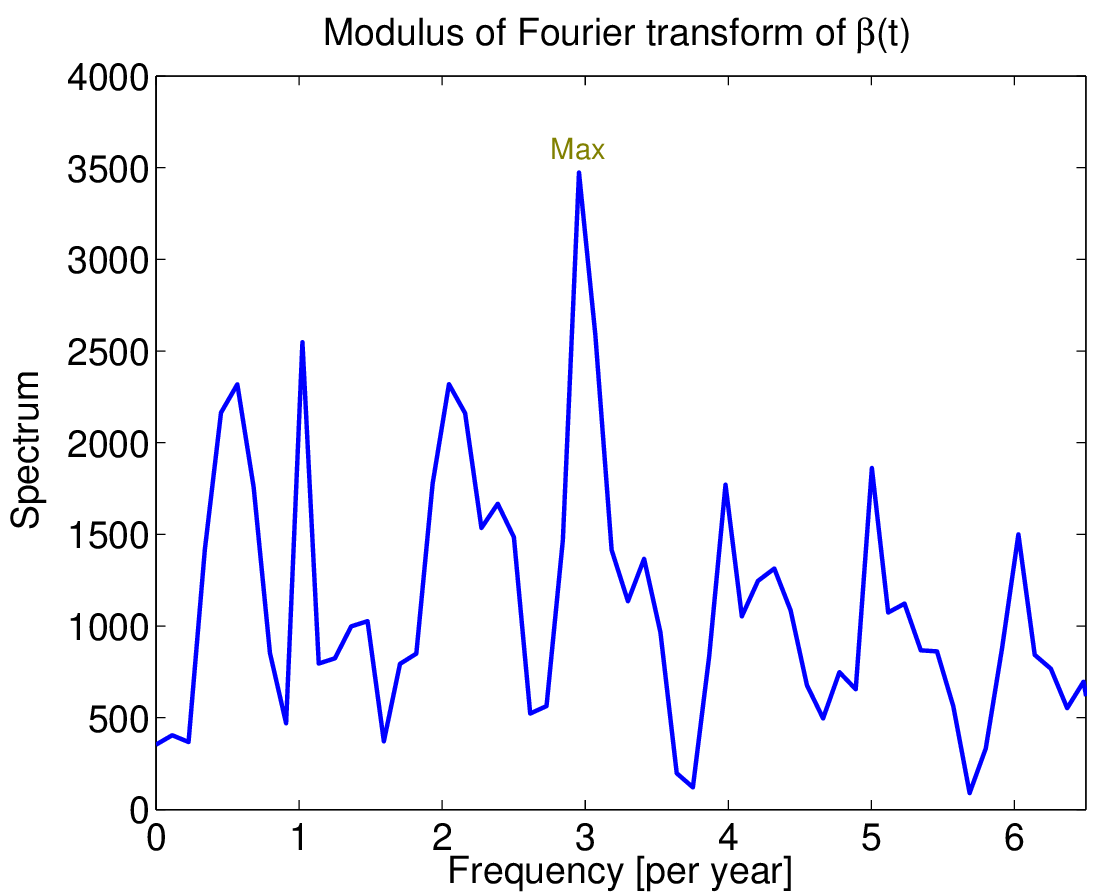}}\\
\subfigure[] {\includegraphics[width=2.5in]{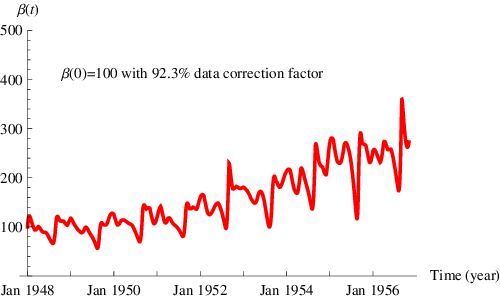}}
\subfigure[] {\includegraphics[width=2in]{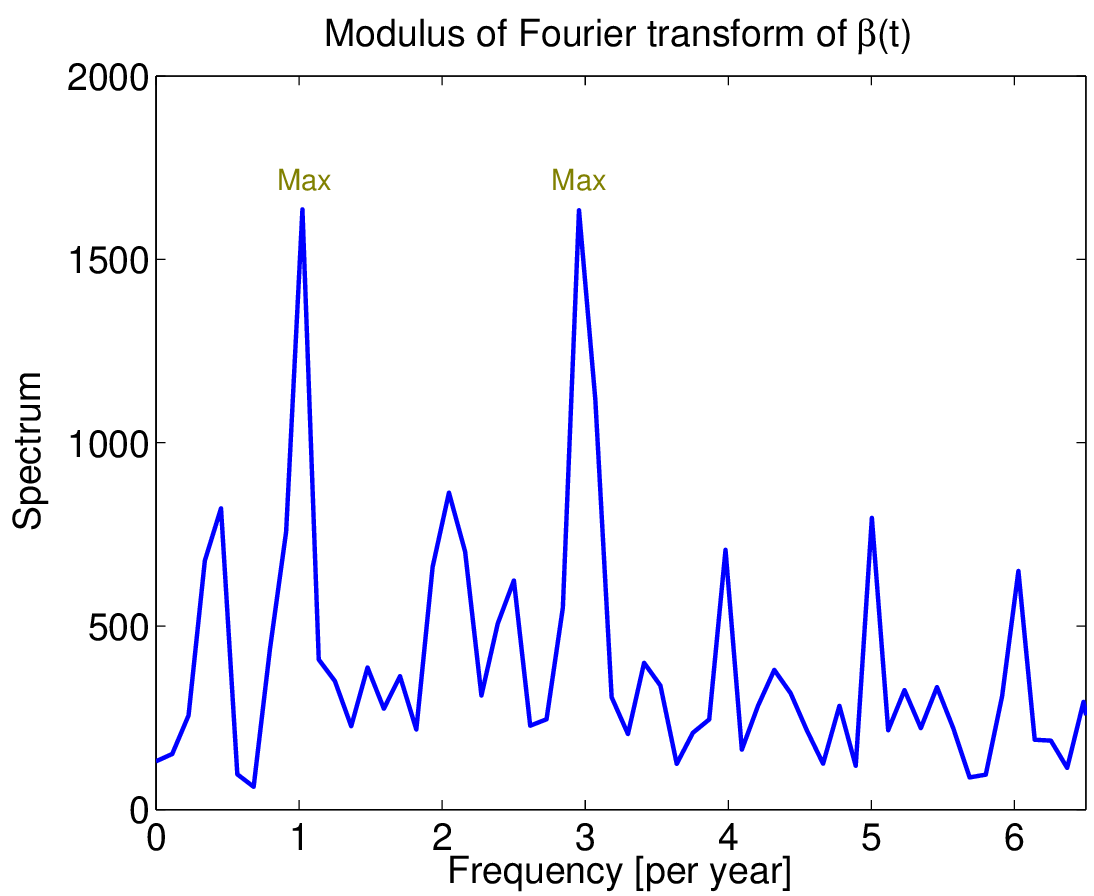}}\\
\subfigure[] {\includegraphics[width=2.5in]{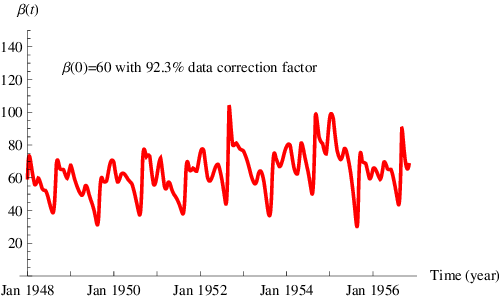}} \subfigure[]
{\includegraphics[width=2in]{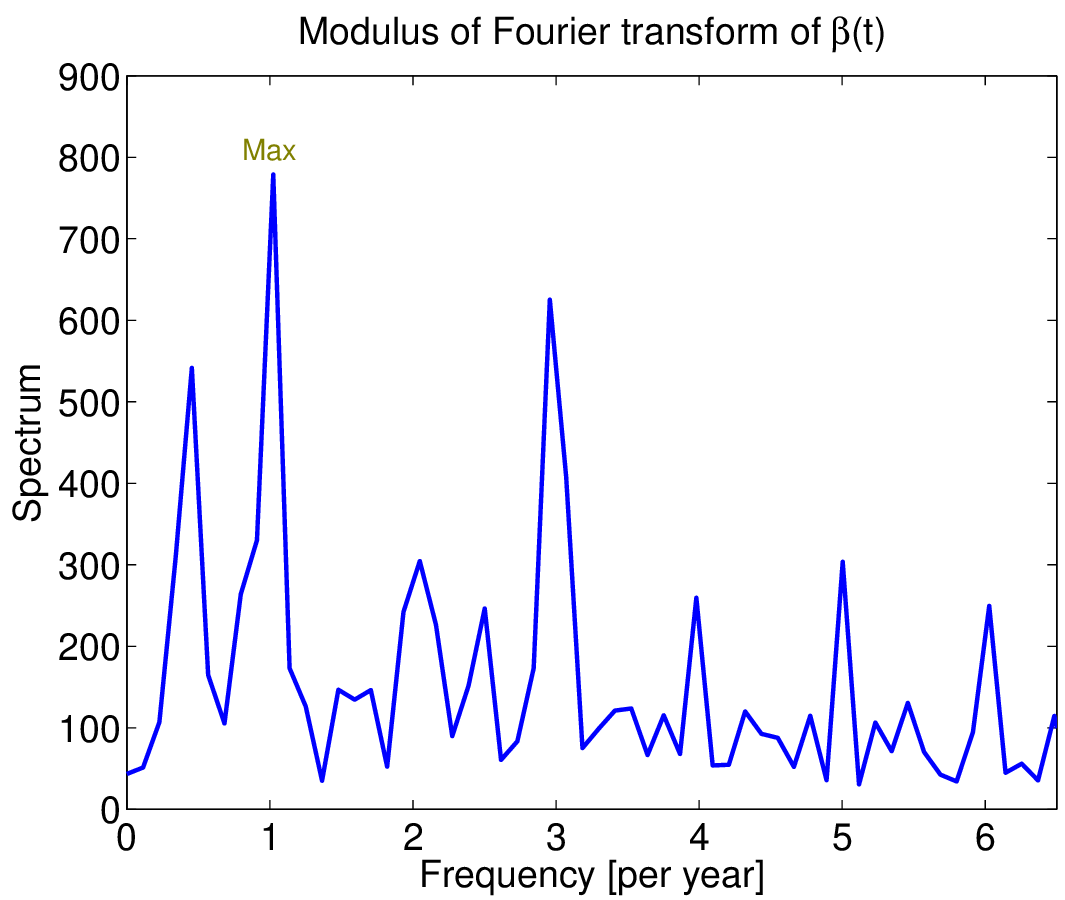}} \\
\subfigure[] {\includegraphics[width=2.5in]{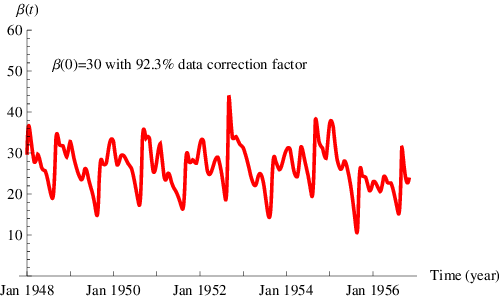}} \subfigure[]
{\includegraphics[width=2in]{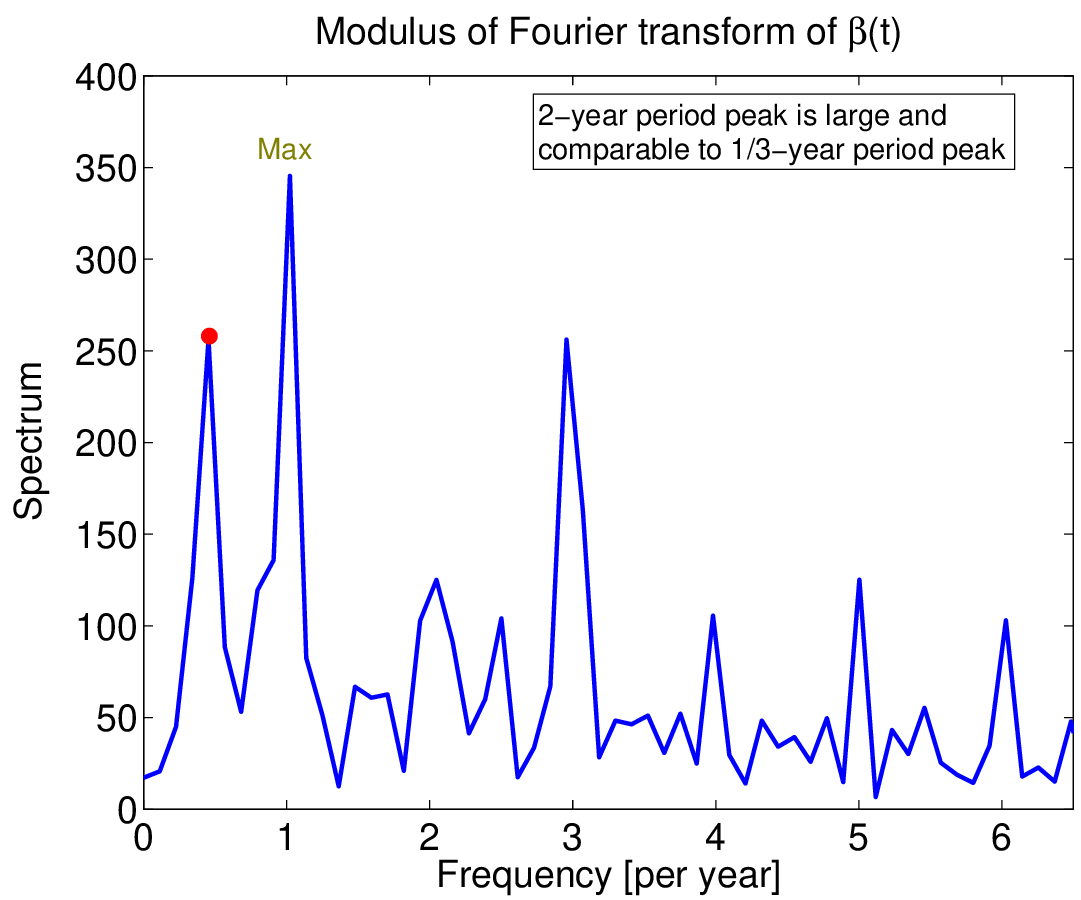}}
\end{array}$
\end{center}
\caption{}\label{fig:beta}
\end{figure}

\begin{figure}[h]
\begin{center}$
\begin{array}{c}
\includegraphics[width=4in]{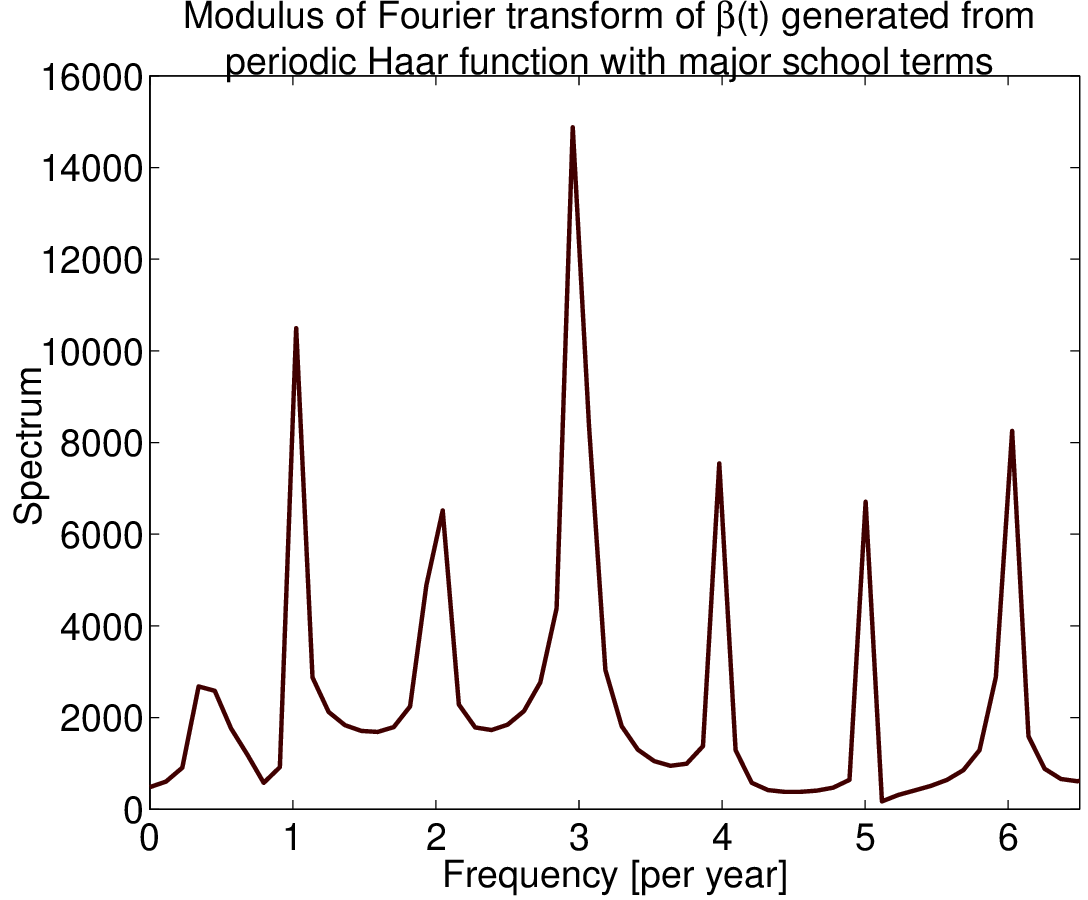}
\end{array}$
\end{center}
\caption{}\label{fig:HaarSpectrum}
\end{figure}

\end{document}